# Nanoparticles modulate contact angle hysteresis in electrowetting

Sumit Kumar[1*], Pawan kumar[2*], SunandoDasGupta[1,3] and Suman Chakraborty[1,2]

[1]Advanced Technology Development Centre, Indian Institute of Technology Kharagpur, India
[2]Department of Mechanical Engineering, Indian Institute of Technology Kharagpur, India
[3]Department of Chemical Engineering, Indian Institute of Technology Kharagpur, India

*Equal Contribution

*Abstract*—The pinning of the contact line adversely influences the electrowetting performance of sessile liquid droplets. In this paper, we report the electrowetting hysteresis characteristics of 100 mM aq. KCl sessile liquid droplet placed on a hydrophobic PDMS surface. The effect of nanoparticles on the contact angle hysteresis under the imposed electric potential is further investigated. This study reveals that the contact angle hysteresis decreases beyond a certain threshold value of nanoparticles concentration. Therefore, nanoparticle suspension in the liquid droplet can be used to enhance or suppress the electrowetting hysteresis and consequentially rate of heat transfer during hot spot cooling.

## I. INTRODUCTION

In last few decades, electrowetting on dielectric (EWOD) has attracted significant attention from the research communities, due to its multifarious applications such as electronic display [1-3], hot-spot cooling [4-6], Lab on chip systems [7-8], variable focal lenses [1-3], auto-focus cell phones [9], video speed smart phones [10-12] and microfluidics devices [13-17]. In electrowetting, the contact angle of a sessile droplet is manipulated using the electrical field. The electrospreading characteristics are adversely influenced by the contact angle hysteresis (CAH), which further deteriorates the performance and the efficiency of the EWOD devices [18-21]. Contact angle hysteresis in EWOD, sometimes called the electrowetting hysteresis, is defined as the difference between advancing contact angle and receding contact angle at the same voltage. The performance of the electrowetting devices depends upon the interactions between the liquid and the surface. The hysteresis is an outcome of the TPCL (three phase contact line) pinning due presence of the physical and chemical heterogeneities on the substrate. The minimum actuation voltage required to spread the sessile liquid droplet drastically increases due to the presence of significant contact angle hysteresis [22-29]. Pinning effects further manifest themselves as stick-slip motion at the TPCL, and affect the switching speed of the EWOD systems [18].

In recent years, several research groups have investigated the CAH behavior and its intricacies during electrowetting. Walker et al. [8] derived a model in which hysteresis was incorporated, but the friction force due to viscous effect is not considered. Later, a new modified Young-Lippmann equation was proposed to take into account the electrowetting hysteresis at low electrical potential [18]. Various groups have also achieved success in minimizing the contact angle hysteresis by using the energy of mechanical vibration. However, this technique does not suit the purpose from the miniaturization perspective [30-34]. Unlike the mechanical shaking, the control of droplet spreading or transport using electrowetting is much higher, which further helps in minimizing the hysteresis and suits for microfluidics applications. The contact angle hysteresis of sessile liquid droplets can be minimized using the alternating electrical potential [35]. However, the reported works available on the influence of nanoparticles in the electrowetting of sessile liquid droplets are very limited. Further, it is of significant importance to investigate the effect of the colloidal particles on contact angle hysteresis of sessile droplets under imposed electrical potential.

In this paper, we report the electrowetting hysteresis characteristics of sessile liquid droplet placed on hydrophobic dielectric thin film. The influence of nanoparticle suspension in liquid droplet on the contact angle hysteresis is further investigated. The experimental findings obtained are demonstrated using the physics of nanoparticles and electrical forces interactions.

## II. MATERIALS AND METHODS

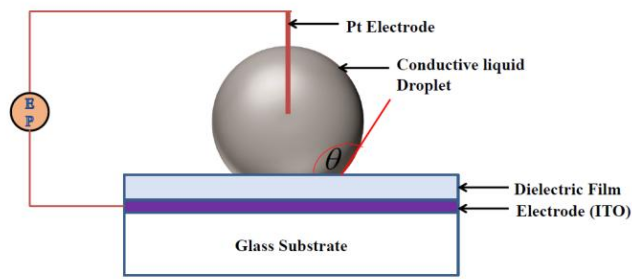

**FIG. 1.** Schematic of standard EWOD setup

The experimental setup is similar as described in Dey et al. [3]. In the present work, the dielectric PDMS film is prepared by mixing the sylgard 180 (silicon based elastomer) with its curing agent in ratio of 10:1 by weight. The PDMS is then coated on transparent indium tin oxide (ITO) glass slides ((Sigma Aldrich, surface resistivity 70–100 Ω/sq.)).

For the coating process, degassed PDMS was poured over an ITO glass slide and then spin coated (using spin coater: Süss MicroTec, Germany), followed by overnight curing at 95 °C. The sessile droplet was prepared from 100 mM potassium chloride (KCl) solution in Milli-Q ultrapure water (Millipore India Pvt. Ltd.). To this aqueous KCl solution, 100 nm polystyrene beads were added in a controlled manner to get 0.025%, 0.05%, 0.075% and 0.1% w/w solutions. The contact angles were measured with the Goniometer (Ramé-hart instrument co., model no. 290-G1).

A sessile droplet of volume $5 \pm 1$ µL was placed on the PDMS coated side of the ITO slide using calibrated microsyringe. A platinum electrode (diameter 160 µm) was then immersed in the droplet (Fig. 1). The connection is completed by grounding the electrode (i.e. ITO slide by removing a small portion of PDMS film from its surface). The value of voltage (V) was progressively increased from 0 V to 200 V with an increment of 10 V using a DC source meter (Keithley 2410, 1100 V SourceMeter). The potential was increased till 200 V and not beyond, to rule out the possibility of contact angle saturation and the dielectric breakdown of the insulation layer on the substrate during the electrowetting. The equilibrium contact angle was measured by using the goniometer and the DROP Image Advanced v2.2.3 software at each applied electric potential value. Once the applied voltage reached 200 volts, the voltage was progressively decreased from 200 V to 0 V with a decrement of 10 V, and the contact angle was again measured at each value of the applied voltage $(V)$. The measurements for each voltage in voltage increasing as well as voltage decreasing mode were done at least five times, and the average value is used. Also the recorded contact angles are the average of left and right contact angles measured by the Goniometer (Ramé-hart instrument co., model no. 290-G1) within an accuracy of $\pm 2^0$. Values of advancing and receding contact angle of the sessile droplet were measured for substrates 10:1 hydrophobic substrate. In addition to the 100 mM aqueous solution, the sessile droplets containing suspended nanoparticles of 100 nm polystyrene latex beads in the 100 mM aqueous solution in the concentration of 0.025, 0.05, 0.075 and 0.1 % w/w respectively, were also used. Effectively, in this experiments a total of 6 different types of sessile droplets (aqueous KCL, 0.025, 0.05, 0.075 and 0.1 % w/w) were used for observing the phenomena of contact angle hysteresis in Electrowetting-on-Dielectrics.

## III. RESULTS AND DISCUSSION

In this section, we present the effect of suspended nanoparticles upon the contact angle hysteresis during electrowetting over the hydrophobic surface. Figure 2 shows the variation of non-dimensional advancing $(\bar{\theta}_{adv})$ and receding $(\bar{\theta}_{rec})$ contact angle for 10:1 hydrophobic PDMS dielectrics, with non-dimensional voltage $(\bar{V} = V/V_{max})$. Addition of nanoparticles to a sessile droplet creates a lubricating effect and reduces the contact line pinning; however, as the concentration of nanoparticles increases, the bulk viscosity of the droplet also increases. Thus, there is competition between both the effects which is evident in figure 2. In the increasing voltage mode, a percentage reduction in contact angle of 17.046 %, 13.762 %, 15.434 %, 18.259 % and 16.614 % was observed for aqueous KCl and 0.025 %, 0.05 %, 0.075 % , 0.1 % w/w nanoparticle concentration sessile droplets (100 nm polystyrene beads) respectively. The sessile droplet with nanoparticle concentration of 0.075 % w/w offered the maximum electro-spreading during electrowetting over 10:1 PDMS surface.

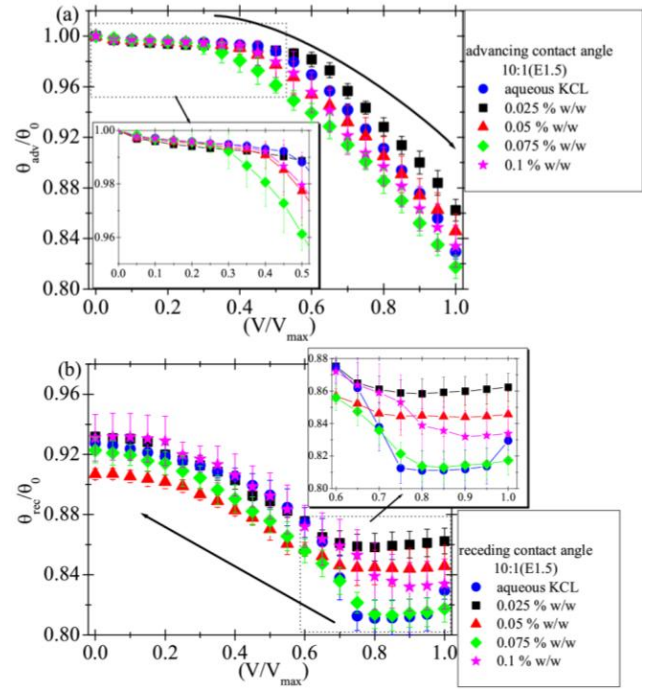

**FIG. 2.** Variation of (a) non-dimensional advancing contact angle $(\bar{\theta}_{adv} = \theta_{adv}/\theta_0)$ and (b) non-dimensional receding contact angle $(\bar{\theta}_{rec} = \theta_{rec}/\theta_0)$ for 10:1 hydrophobic PDMS dielectrics with non-dimensional voltage $(\bar{V} = V/V_{max})$ and sessile droplets containing suspended nanoparticles in 0.025, 0.05, 0.075 and 0.1 % w/w respectively.

Figure 3 shows the variation non-dimensional contact angle hysteresis $(\bar{H})$ with $\bar{V}$ for electrowetting over a rigid dielectric surface (10:1 PDMS). Addition of nanoparticles to the sessile droplet was observed to increase the contact angle hysteresis $(\bar{H})$ (fig. 3). The nanoparticles arrange themselves in a solid like ordering [36-39] at the TPCL and causes the pinning of the contact line. During the reduction of voltage after reaching $V_{max}$, the capillary forces pull the contact line back to corresponding value at the new reduced voltage against the push of electric force component [1, 4] and the pinning force at the TPCL. An increased pinning force upon addition of nanoparticles requires a larger reduction in the local contact angle $(\theta_{local})$ before it becomes equal to the electrowetting receding angle $(\theta_{local,rec})$ [18] and thus a greater contact angle hysteresis (see fig. 3). However, at the nanoparticle concentration of 0.1 % w/w, a reduction in $\bar{H}$ was observed (fig. 3). At higher nanoparticle concentration (0.1 % w/w), a reduction in contact line pinning was observed during the decreasing voltage mode (fig. 2 (b)) and consequently a much lesser contact angle hysteresis was observed for 0.1 % w/w nanoparticle concentration droplet (fig. 3 ). Another important observation from figure 2 is that that the maximum contact angle hysteresis $(\bar{H}_{max})$ does not occur at the zero voltage, but at a finite voltage. The reason for this apparent anomaly is that, due to the pinning of the contact line, the contact angle starts decreasing only when the applied voltage becomes equal to the actuation voltage $(\bar{V}_{act})$ for the system.

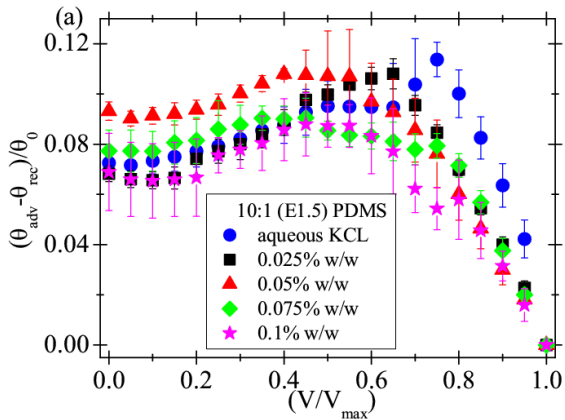

**FIG. 3.** Variation of non-dimensional contact angle hysteresis $(\bar{H}=(\theta_{adv}-\theta_{rec})/\theta_0)$ with non-dimensional voltage $(\bar{V}=V/V_{max})$ and sessile droplets containing suspended nanoparticles in 0.025, 0.05, 0.075 and 0.1 % w/w respectively for 10:1 hydrophobic PDMS surface.

## IV. CONCLUSIONS

In this paper, the influences of colloidal suspension and electrical potential on contact angle hysteresis of sessile liquid droplets have been explored. This study unveils a reduction in contact angle hysteresis at higher concentration of nanoparticles. In addition, it is also observed that the maximum hysteresis occurs at a finite electrical potential (not at zero potential). The power to alter the spreading of droplets using colloidal suspension along with electrical potential may find its applications in several heat transfer and engineering fields such as electronics cooling, liquid lenses, biological assay and DMF based medical diagnosis.